\documentclass[a4paper,11pt]{article}
\usepackage{pos}
\usepackage[numbers]{natbib}

\usepackage{hyperref}

\usepackage{amsbsy}
\usepackage{comment}
\newcommand{\dd}{\text{d}}

\title{Quantum causality in kappa Minkoswki}

\author*[a,b]{Valentine Maris}

\affiliation[a]{IJCLab, Université Paris Saclay, \\
  91405 Orsay, France}
\affiliation[b]{LPENSL, ENS de Lyon\\
  69007 Lyon, France}
\emailAdd{valentine.maris@ens-lyon.fr}

\abstract{Recent results on causality in noncommutative space-time are reviewed. We study, in particular, quantum causal structures in 1+1 dimensional kappa Minkowski space-time. This later is described by a twisted Lorentzian Spectral Triple build with a twisted set of derivatives. Investigation of causality provides a quantum constraint, which is a quantum analog of the speed light limits.}

\FullConference{Proceedings of the Corfu Summer Institute 2024 "School and Workshops on Elementary Particle Physics and Gravity" (CORFU2024)\
12 - 26 May, and 25 August - 27 September, 2024\\
Corfu, Greece\\}

\renewcommand{\logo}{\relax}

\begin{document}
\maketitle

\section{Introduction}
Causality is a keystone of Quantum Field Theory, General Relativity, and any serious and realistic physical theory.
However, in Quantum Gravity, the description of the causal structure is not fully achieved.  In particular, by allowing gravity to have quantum properties, the superposition principle would lead to a superposition of time-like and space-like intervals \cite{brukner2014quantum}. Thus, the nature of separation, causal (i.e., time-like) or not causal (i.e., space-like), between two events may not be fixed. This could lead to inconsistencies and highlights that the notion of causality in gravity must be reconsidered at the quantum level. \\

Signs of causal deformation would greatly advance our understanding of quantum gravity, giving new constraints on models; for a phenomenological approach, see \cite{Addazi_2022} for a review including possible causality tests at the Planck scale. \\
Several attempts have been made in the literature to define a proper causal structure for quantum gravity. One of the first approaches was with the so-called "Causal sets" \cite{PhysRevLett.59.521}, a discretized version of space-time with the usual notion of causality on it. 
Focusing on Lorentzian noncommutative geometry, deformed or fuzzy causal structures have been investigated. Remember that any notion of causal structure on quantum space-time should, at the commutative limit, give back the usual notion of causality. Fuzziness and Localization of events living in different frames have been studied in \cite{Lizzi_2019} and \cite{Lizzi:2022hcq} (for various kinds of spaces).
In \cite{Mercati:2018ruw} \cite{Mercati:2018hlc}, the light cone structure is examined; this later has blurred contours, with "blurriness thickness"  of the size of Planck length. \\

Here, we will focus our attention on an algebraic formulation of causality on the Lorentzian Spectral Triple. Those structures are a Lorentzian adaptation of the Riemannian Spectral Triple of Alain Connes \cite{connes1994noncommutative}.
An important observation in this framework is the lack of points. Therefore, the usual notion of time-like or space-like distance between events (i.e., points) can not be applied. Two approaches have been developed to bypass this problem; the first one \cite{Besnard_2009} is based on iscone.
The second one in \cite{Franco_2013} proposes an algebraic version of causality, which is shown to be equivalent to the usual one at the commutative limit, and this formalism is later applied in \cite{franco2014noncommutativegeometrylorentzianstructures}  to Lorentzian Spectral Triple. In those structures, the Dirac operator plays a central role, acting as a metric and thus allowing to define a notion of quantum causality. Those approaches of algebraic causality have been explored in both almost commutative manifold \cite{Franco_2014} and Moyal space \cite{Franco_2016}.\\

Here, we will explore the fate of causality on a very promising noncommutative space-time: $\kappa$ Minkowski \cite{Lukierski:1992dt}. This space can be viewed as quantum deformations of Minkowski space-time and has been gaining huge interest for years. It is defined \cite{Majid_1994} as the dual of a subalgebra (composed of deformed translations) of $\kappa$ Poincaré algebra with coordinates satisfying $[x_0,x_j] = \frac{i}{\kappa}x_j, \; [x_i,x_j] = 0$. Here, $\kappa > 0$ is the deformation parameter encoding quantum gravity effects. At the commutative limit, space-time and symmetry space return the usual Minkowski and Poincaré spaces. We will work with a $(1 + 1)$ dimensional version of $\kappa$ Minkowski, as there are no difficulties adapting the work to $(1+3)$ dimensions. Constructing a related Lorentzian Spectral Triple is not easy or unique. Different choices for the Dirac operators can be made and lead to different results regarding the causal structure of the space.\\
A first exploration of causality in the framework of Lorentzian Spectral Triple on $\kappa$ Minkowski has been done in \cite{Franco_2022}. The result of this study was a constraint view as an analog of the classical speed of light. A huge drawback of this model is that at the commutative limit, one derivative vanishes, which results in an unphysical situation. Here, we will focus on the second attempt made in \cite{Franco_2023}, where we build a suitable Dirac operator coinciding at the commutative limit with the usual one. However, this operator is built with a twisted derivative, i.e., satisfying a twisted form of Leibniz rule \footnote{at the commutative limit we find back the usual Leibniz rule}, given thus, a twisted version of Lorentzian Spectral Triple. \\

The first part will remind us of causality in physics and how to end up with an equivalent and algebraic version. Then, we will detail how to extend algebraic causality to noncommutative space. $\kappa$-Minkowski will be introduced in a second part, and the construction of the Lorentzian will be explained. The last part will study the notion of causality for the Lorentzian Spectral Triple on $\kappa$-Minkowski. Finally, we will exhibit quantum constraints and interpret them.

\section{Algebraic formulation of causality in Noncommutative geometry}

The usual formulation of causality uses points on the manifold (i.e., events) and causal curves connecting them. However, in noncommutative geometry, there is no notion of points. Thus, the usual definition of causality can not be applied. Franco and Eckstein find a way to bypass this problem by finding a new and equivalent causality formulation that does not rely on points \cite{Franco_2013} and is shown to be compatible with noncommutative geometry.

\subsection{Causal evolution }

Let $\mathcal{M}$ be a Lorentzian manifold equipped with metric $g$, and  $A,B \in \mathcal{M}$ two points of space-time. The relation "$ B$ is in the causal future of $A$",  is verify if:
\begin{equation}
    \begin{split}
        \exists c: \mathbb{R} \to \mathcal{M}, \;  & \exists t_1 \leq t_2 \in \mathbb{R},  \quad c(t_1) = A, \; c(t_2) = B \\
        & \forall t \in [t_1, t_2], \quad  g(c'(t),c'(t)) \leq 0
        \label{condition1}
    \end{split}
\end{equation}
This condition indicates the existence of a smooth curve connecting $A$ to $B$, which is future-directed at any instant. By future-directed, we mean that the tangent vectors are timeline and pointing in a "future direction" depending on a choice of time arrow. 

\begin{figure}[ht]
    \centering
   \includegraphics[width=0.4\textwidth]{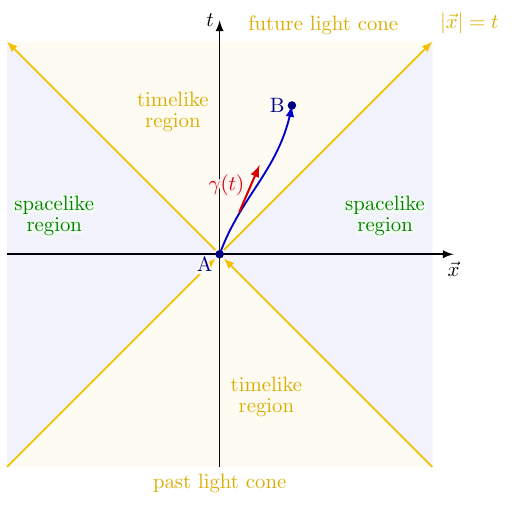}
   \caption{Two events $x$ and $y$ are causally related by a curve with only future time-directed tangent vectors. One can describe a light cone structure for every event.  }
\end{figure}

However, noncommutative geometry lacks manifold, points, metric, and tangent vectors; therefore, such formalism of causal structure is not adapted. To solve this problem, we will use a different formulation of causality.  The first step is to express causal relations between points with functions instead of tangent vectors. It has been shown, \cite{Franco_2013}, that the causal structure of the Manifold is entirely determined by the set of the causal functions $f: \mathcal{M} \to \mathbb{R}$, which are non-decreasing along every future-directed curve. Then, one has: \\

\noindent
\textbf{Proposition [Algebraic causality]} \\
\textit{Let $(M,g)$ be a globally hyperbolic Lorentzian manifold and $\mathcal{C}(M)$ be the set of smooth bounded causal functions. Then, the causal structure of the manifold is completely determined by:}
 \begin{equation}
    \forall x,y \in M, \quad x \preceq y \longleftrightarrow \forall f \in \mathcal{C}(M), \; f(x) \leq f(y)
\end{equation}
and this proposition is equivalent to the usual notion of causality \cite{Franco_2013}. The next step will be to find the equivalent of a point in noncommutative geometry and apply this algebraic causal relation.  

\subsection{Noncommutative geometry and Gelfand theory }

Manifolds are equivalent to commutative $\mathbb{C}^{*}-$algebra in commutative geometry. Taking the set of continuous functions, $\mathbb{C}$-valued, with the usual point-wise product. This form a commutative Banach $C^{*}-$algebra, denoted by $\mathcal{A} = \mathcal{C}_0(\mathcal{M})$. Considering the spectrum $\Delta(\mathcal{A})$ composed of all non-zero $*$-homomorphisms from $\mathcal{A}$ to $\mathbb{C}$. Its elements are called characters. The Gelfand Naimark theorem gives an isomorphism between the spectrum of the commutative $C^{*}-$algebra and the manifold $\mathcal{M}$. So there is the same amount of information in the manifold as in the $C^{*}-$algebra. Going to the noncommutative point of view, one can take a noncommutative $C^{*}-$algebra. The spectrum of $\mathcal{A}$ is now composed of pure states\\

\noindent
\textbf{Definition [Pure state]}\\
\noindent
\textit{For $\mathcal{A}$ a $C^{*}$-algebra, a pure state is a positive linear function of norm one that cannot be written as a convex combination of two other states. We will denote $\mathcal{P}(\mathcal{A})$ the set of pure states .}\\

Those pure states are for a noncommutative algebra, the equivalent of points for the manifold. At this stage, we have a new formulation of causality and a notion of "noncommutative points." We will now generalize the differential structure given by a manifold and a metric using Spectral Triple.

\subsection{Lorentzian Spectral triple}\label{sectionLST}

The Riemannian Spectral Triple was initially introduced by Alain Connes in \cite{connes1994noncommutative}. The central piece of this structure is the Dirac operator, which contains all information about the metric and the geometry. A Lorentzian version of Spectral Triple has been developed by Franco in  \cite{franco2014noncommutativegeometrylorentzianstructures}. It is defined by the following set of objects:
\begin{equation}
    \big\{\mathcal{A},\ \widetilde{\mathcal{A}},\ \pi,\ \mathcal{H},\ D,\ \mathcal{J}  \big\},
\end{equation}
where $\mathcal{A}$ is an involutive non-unital pre-$C^{*}$-algebra, $\widetilde{\mathcal{A}}$ a preferred unitalization of $\mathcal{A}$ \footnote{Unitalization is necessary for technical purpose. Note that one cannot only consider an unital $\mathcal{C}^{*}$-algebra, which is a compact space where the notion of no notion of causality can be defined.}. $\mathcal{H}$ is an Hilbert space. $\pi : \mathcal{A} \text{ or } \widetilde{\mathcal{A}} \to \mathcal{B}(\mathcal{H})$ a faithful $*$-representation into the algebra of bounded operators on $\mathcal{H}$. $ \mathcal{D}$ the (unbounded) Dirac operator with domain dense in $\mathcal{H}$ verifying 
\begin{equation}
    \forall a \in \widetilde{\mathcal{A}}, \quad [ \mathcal{D},\pi(a)] \in \mathcal{B}(\mathcal{H}).
\end{equation}
$\mathcal{J} \in \mathcal{B}(\mathcal{H})$ the fundamental symmetry such that,
\begin{equation}
    \mathcal{J}^{2} = 1, \; \mathcal{J}^\dagger = \mathcal{J},\; \forall a \in \widetilde{\mathcal{A}}, \; [\mathcal{J}, \pi(a)] =0, \; \mathcal{D}^\dagger \mathcal{J} = - \mathcal{J} \mathcal{D}.
\end{equation}
$\mathcal{J}$ act on the Hilbert product by turning it into an indefinite Krein product,
\begin{equation}
    (.\ ,\ .)_\mathcal{J}=\langle\ .\ , \mathcal{J}\ .\rangle.
\end{equation}
This fundamental symmetry plays the role of a Wick rotation here. Finally, let us introduce $\mathcal{T}$ a self-adjoint operator and $\mathcal{N}$ a positive operator of $\widetilde{\mathcal{A}}$ with dense domain in $\mathcal{H}$. They should verify the following conditions:
\begin{equation}
    \mathcal{J} = -\mathcal{N}[D, \mathcal{T}] \quad (1 + \mathcal{T})^{-\frac{1}{2}}\in \widetilde{\mathcal{A}}.
\end{equation}
$\mathcal{T}$ is viewed as a noncommutative generalization of time function. Changing $\mathcal{T} \to - \mathcal{T}$ (i.e., doing time reversal) allows to construct a well-defined Lorentzian Triple, with $\mathcal{J} \to - \mathcal{J}$ satisfying all axioms $(3)-(7)$. The left condition ensures a Lorentzian-type signature. 

\subsection{Causality in noncommutative geometry}

Having defined the Lorentzian Spectral Triple, we will put on it a causal structure thanks to the algebraic formulation of Franco and Eckstein \cite{Franco_2013}. First, let us introduce a causal order on $\mathcal{P}(\widetilde{\mathcal{A}}$ the space of states.\\ 

\noindent
\textbf{Definition [Causal cone]} \\
\noindent 
\textit{A causal cone $\mathcal{C} \in \widetilde{\mathcal{A}}$ is define as, } $\forall a,b \in \widetilde{\mathcal{A}},\; \forall \lambda \in \mathbb{R}_+, \; \forall x \in \mathbb{R}, \; \forall \phi \in \mathcal{H}$
\begin{equation}
    \begin{split}
         a^\dagger = a, \; a + b & s\in \mathcal{C},  \; \lambda a \in \mathcal{C}, \; 1 x \in \mathcal{C}, \, \overline{\text{span}_\mathbb{C}} = \overline{\widetilde{\mathcal{A}}} \\
        & \langle \phi, \mathcal{J} [D, \pi(a)] \phi \rangle \leq 0
        \label{causalcone}
    \end{split}
\end{equation}
\textit{The last condition is the noncommutative analogue of (\ref{condition1}) on tangent vectors. The sign of \ref{causalcone} is a choice and can be changed by $\mathcal{J} \to - \mathcal{J}$} \\

The causal evolution between pure states is given by,
\begin{equation}
    \forall \omega, \eta \in \mathcal{P}(\widetilde{\mathcal{A}}) \quad \omega  \preceq  \eta \Longleftrightarrow \forall a \in \mathcal{C}, \; \omega(a) \leq \eta(a)
    \label{causalevolution}
\end{equation}
This evolution is well-defined since states are positive linear functional. Furthermore, the relation $ \preceq $ is a partial order of the space of states (see Proposition 6,  \cite{Franco_2013}) and thus defines an algebraic version of the usual notion of causality. This concludes our description of causal structures in noncommutative spaces; we will now apply it to a particular space.

\section{Kappa-Minkowski}

A very promising family of noncommutative spaces that emerged last year is a deformed version of Minkowski space. Historically, the Hopf algebra $\kappa$-Poincaré was introduced in \cite{Lukierski:1992dt}. Then Majid and Ruegg build $\kappa$-Minkowski as the Hopf dual of the translational part of $\kappa$-Poincaré \cite{Majid_1994}. The phenomenology implications of such a deformed model are numerous, especially in the theory of quantum space-time \cite{ADDAZI2022103948}, which explains the heavy interest in $\kappa$-deformation. Here, $\kappa$ is understood as a deformation parameter associated with Planck length. \\

In this section, we will briefly recall some basic facts about $\kappa$-Minkowski and $\kappa$-Poincaré, then build the associated Lorentzian Spectral Triple and finally study causal structure.

\subsection{Deformed space-time and its associated Hopf algebra, kappa Poincaré}

$\kappa$-Minkowski space can be understood as the universal enveloping algebra of the Lie coordinates with commutator $[x_0,x_j] = \frac{i}{\kappa} x_j$.To avoid complicated computations,  we will restrict our analysis to the $(1+1)$ dimensional case, the higher dimensional situation being similar. The associated Lie group is the affine group of the real line \cite{khalil1974analyse} \cite{gelfand1947unitary} $\mathcal{G} = \mathbb{R} \ltimes \mathbb{R}$. Taking Haar measure, which reduces here to Lebesgue, one can build the convolution algebra $(L^{1}(\mathcal{G}), o, \dagger)$ with convolution product $o$ and involution $\dagger$. The star product is obtained from  Weyl quantization map \footnote{Weyl quantization map is defined by $ Q(f) : = \pi \mathcal{F}(f)$, for any $f \in L^{1}(\mathcal{G})$, $\pi$ a non-degenerate star representation. In order to find $f \star g$, one must remember that $\pi$ is non-degenerate and a $*$-morphism of algebra.}\cite{Durhuus_2013} \cite{Poulain_2018}, with $f \star g := \mathcal{F}^{-1}(\mathcal{F}(f) o \mathcal{F}(g))$ and reads:
\begin{equation}
    \begin{split}
        (f\star g)(x_0,x_1) & =\int \frac{ \dd p_0}{2\pi}\dd y_0\ e^{-iy_0p_0}f(x_0+y_0,x_1)g(x_0,e^{-p_0/\kappa}x_1) \\
        f^\dag(x_0,x_1)&= \int \frac{\dd p_0}{2\pi}\dd y_0\ e^{-iy_0p_0}{\bar{f}}(x_0+y_0,e^{-p_0/\kappa}x_1)
    \end{split}
\end{equation}
With $f,g \in \mathcal{A}_x = (\mathfrak{C}(\mathbb{R}, \mathfrak{C}(\mathbb{R}),\star,\dagger)])$ the space of functions whose analytic continuation in the first variable is an entire function on $\mathbb{C}$ of exponential type with values in the space of analytic
functions of exponential type in the second variable. With a Fourier transform, one can use another set of variables, $(p_0,p_1)$ and functions on $\mathcal{A}_p = \mathcal{C}^\infty_{c}(\mathbb{R},\mathcal{C}^\infty_{c}(\mathbb{R})$. \\

The symmetry space of $\kappa$-Minkowski is given by the Hopf algebra $\kappa$-Poincaré generated with translations and a boost $\{P_0,P_1,N \}$. Those generators satisfy commutation relations and Hopf algebra structure equations we will not detail here but can be found in the appendix of \cite{Franco_2022}. To build the Dirac operator, we need to define derivatives, 
\begin{equation}
    X_0 = (1 - \mathcal{E}), \quad X_1 = P_1 \text{ with } \quad \mathcal{E} := e^{-P_0/\kappa} 
    \label{kappaDer}
\end{equation}
Two remarks are needed. First, we retrieve the usual derivatives $-i\partial_0, \; - i \partial_1$ at the commutative limit. This was not the case in the previous study on causality in $\kappa$-Minkowski by \cite{Franco_2022}, where one derivative vanishes at the commutative limit. Secondly, the price to pay is that now the derivatives are said to be "twisted", meaning they satisfy a twisted Leibniz rule,
\begin{equation}
    X_\mu(f \star g) = X_\mu(f) \star g + (\mathcal{E} \triangleright f) \star X_\mu(g)
    \label{twistedLeibniz}
\end{equation}
where $\mathcal{E}$ is an automorphism of of $\kappa$-Minkowski and $\triangleright$ the right action of $\mathcal{E}$ on $f$. At the commutative limit $\mathcal{E} \to \mathbb{I}$ and the usual Leibniz rule is recovered.

\subsection{A Lorentzian Spectral Triple for kappa Minkowski}

Using formalism describe in \ref{sectionLST} one can build a suitable Lorentzian Spectral Triple for $\kappa$-Minkowski. Particular attention is given to derivatives choice (which gives the Dirac operator) and representations of $\kappa-$Minkowski. Unitary representations of the affine group $\mathcal{G}$ have been classified by \cite{khalil1974analyse} \cite{gelfand1947unitary}. The non-trivial is given by
\begin{equation}
    \begin{split}
        \widetilde{\pi}_\pm:& \mathcal{G} \to \mathcal{B}(L^{2}(\mathbb{R},ds)) \\
       &  (p_0,p_1) \to \widetilde{\pi}_{\pm}(p_0,p_1) \quad  \text{with} \quad  (\widetilde{\pi}_{\pm}(p_0,p_1)\phi)(s) = e^{\pm i p_1 e^{-s}}\phi(s +p_0)
    \end{split}
\end{equation}
with $\phi \in L^{2}(\mathbb{R},\dd s)$, $\dd s$ the usual Lebesgue measure. Other unitary and irreducible representations of the affine group are trivial. The non-degenerate, irreducible, and bounded $*$-representations of the convolution algebra stem from the one above and read:
\begin{equation}
    \begin{split}
         &\pi_\pm:\mathcal{A}_p \to \mathcal{B}(L^2(\mathbb{R},\dd s)) \\
         &(\pi_\pm(f)\phi)(s)=\int \dd^{2}p\ e^{p_0}\mathcal{F}f(p_0,p_1)e^{ \pm i p_1e^{-s}}\phi(p_0+s) 
         \label{repreP}
    \end{split}
\end{equation} 
Doing a Fourier transform, one can find the associated (non-degenerate, irreducible and bounded) $*$-representations of the $*$-algebra,
\begin{equation}
    (\pi_\pm(f)\phi)(s)= \int \dd u \dd v  f(v, \nu e^{-s})e^{\mp i(u-s)} \phi(u)
    \label{repreX}
\end{equation}
The Hilbert space of our Lorentzian Spectral Triple is a direct sum of $\mathcal{H}= (\mathcal{H}_+ \oplus \mathcal{H}_-) \otimes \mathbb{C}^{2}$, with $\mathcal{H}_\pm:= L^{2}(\mathbb{R},\dd s)$ and the associated faithful representation $\pi = (\pi_+ \oplus \pi_-)\otimes \mathbb{I}_2)$. The tensorial product with $\mathbb{C}^{2}$ in the definition of the Hilbert space is due to the fact that we consider a two-dimensional Dirac operator\footnote{The case with full $4$ d $\kappa$-Minkowski space is similar by enlarging the Hilbert space to $\mathcal{H}= -(\mathcal{H}_+ \oplus \mathcal{H}_-) \otimes \mathbb{C}^{4}$ }. The scalar product of $\mathcal{H}$ is given by $\langle \Phi, \Psi \rangle = \sum_{\nu = +,-} \langle \phi^{(\nu)},\psi^{(\nu)} \rangle_{\mathcal{H}_\nu}$ with $\langle .,. \rangle_{\mathcal{H}_\pm}$ the usual scalar product of $L^{2}(\mathbb{R},\dd s)$. However, the Lorentzian spectral triple needs to have a Krein product. This requires the definition of the fundamental symmetry and the time operators. 
\begin{equation}
    \mathcal{J} = i \gamma^{0} \otimes \mathbb{I}_2 \quad \mathcal{T} = - i \oplus_\nu (\pi_\nu(x_0) \otimes \mathbb{I}_2)
\end{equation}
One can check that all the operators within the faithful representation and the Hilbert space verify the conditions of a Lorentzian Spectral Triple.
The Krein product can now be defined as:
\begin{equation}
    \langle \Phi, \Psi \rangle_{\mathcal{J}} = \langle \Phi, \mathcal{J} \Psi \rangle = \sum_{\nu = +,-} \langle \phi^{(\nu)}, i \gamma_0 \; \psi^{(\nu)} \rangle_{\mathcal{H}_\nu}
\end{equation}
Another important ingredient of the Lorentzian Spectral Triple is the Dirac operator. It encodes all the differential structures and metrics of a manifold algebraically. This later is defined with twisted derivatives \ref{kappaDer}
\begin{equation}
    \mathcal{D} = -i \gamma^\mu X_\mu \otimes \mathbb{I}_2 = \begin{bmatrix}
        0 & X_- \\
        X_+ & 0
    \end{bmatrix} \otimes \mathbb{I}_2:= D \otimes \mathbb{I}_2, \quad X_\pm = X_0 \pm i X_1
    \label{dirac}
\end{equation}
This operator does not fulfill all conditions $(4)-(7)$, but a twisted version. 
Euclidean twisted spectral triple has been define in  \cite{connes2006typeiiispectraltriples}
and adapted for $\kappa$-Minkowski in \cite{Matassa_2014}. A Lorentzian version has been developed in \cite{Franco_2023}. The main difference with the usual Spectral triple one is that the bracket must be twisted to remain bounded: 
\begin{equation}
    \forall a \in \widetilde{\mathcal{A}}, \quad [\mathcal{D}, \pi(a)]_{\mathcal{E}} \in \mathcal{B}(\mathcal{H})
\end{equation}
Other twisted conditions are given by,
\begin{equation}
    \mathcal{D}^\dagger = - \mathcal{J} \rho^{-1} \mathcal{D} \rho \mathcal{J}, \quad \rho := \mathcal{E} \otimes \mathbb{I}_2, \quad [\mathcal{\mathcal{D},\mathcal{T}}]_{\mathcal{E}} = - \mathcal{N} \mathcal{T}
\end{equation}

\section{Quantum constraints on kappa Minkowski}

Using the Lorentzian Spectral triple defined in the previous section, we can now apply the notion of algebraic causality of section 1. First, one needs pure states, i.e., the equivalent of points for noncommutative $*$-algebra. The complete set of pure state of $\kappa$-Minkowski, $\mathcal{P}(\mathcal{M}_\kappa)$, is not fully determined but is non zero and some explicit families of such state are well known. Let consider 
\begin{equation}
    \omega\pm^{\Phi}: \mathcal{A} \to \mathbb{C}, \quad \omega^{\Phi}_\pm(a) = \langle \Phi, \pi_\pm(a) \Phi \rangle, \quad \Phi \in \mathcal{H}_\pm, \quad ||\Phi || =1  
    \label{pureS}
\end{equation}
This state is pure as they are cyclic vectors of the form $\langle v, \phi(a) v \rangle$ for $a \in \mathcal{A}$, $v \in \mathcal{H}$, for an irreducible representation $(\pi,\mathcal{H})$. Those  pure states can be extended to $\widetilde{\mathcal{A}}$ restricting $\Phi \in \mathcal{H}_\pm$ to smooth compactly support functions in $\mathcal{H}_\pm$\\
Having pure states, we are looking for functions in the causal cone to apply causal evolution \ref{causalevolution} between pure states. The casual cone \ref{causalcone} associated with the Dirac operator \ref{dirac} gives a condition that every function of the causal cone $\mathcal{C}$ should verify:
\begin{equation}
    \forall a \in \mathcal{C},\quad 
     \langle \phi, \pi_\pm(X_\pm(a)) \phi \rangle_{\mathcal{J}} \geq 0, 
     \label{causalconeexplicit}
\end{equation}
Using the explicit expression of the representation \ref{repreX}, one gets: 
\begin{equation}
    - i \int \dd s \dd u (1 - e^{s-u})a(u-s, \nu e^{-s}) \mp \partial_\beta a(u-s,\nu e^{-s})\phi(u) \overline{\phi}(s) \leq 0
\end{equation}
This inequality can be solved for particular functions, showing that the causal cone is non-empty. The first evident solution is given by $T = x_0$, the time function. A second one is the class of "Lambda light cone coordinates" functions $a_\lambda(x_0,x_1) = x_0 \pm \lambda x_1$ with $\lambda \in [-1, 1]$. Such solutions are reminiscent of the light cone structure. A last family of solutions can be found by using the "split function" of the form $a(x_0,x_1) := h(x_0) + g(x_1)$ with $i X_0(h) = 1$ and $|g'(\nu e^{-s})| \leq 1$. Taking the non-empty causal cone, one can write the causal relation between two pure states:
\begin{equation}
\begin{split}
     & \omega_\pm^{\phi_1} \preceq \omega\pm^{\phi_2}\Longleftrightarrow \forall a \in \mathcal{C}_x, \;  \omega_\pm^{\Phi_1}(a) \leq \omega_\pm^{\Phi_2}(a) \\
    &\int \dd s \dd u a(u-s, \pm e^{-s})[\overline{\phi_2}(s) \phi_2(u) - \overline{\phi_1}(s) \phi_1(u)] \geq 0
\end{split}
\end{equation}
With the a particular solution, $a(x_0,x_1) = x_0 \pm x_1$, one can derive the following quantum constraint:
\begin{equation}
    \langle \phi_2 | P | \phi_2 \rangle - \langle \phi_1 | P | \phi_1 \rangle \geq |\langle \phi_2 | X | \phi_2 \rangle - \langle \phi_1 | X | \phi_1 \rangle|
\end{equation}
Denoting the variation of the expectation value by $\delta \langle A \rangle $, we get
\begin{equation}
    \delta \langle P \rangle \geq |\delta \langle X \rangle|
    \label{average}
\end{equation}
This constraint can be interpreted as a quantum analogous to the classical speed of light limit 
\begin{equation}
    \delta t \geq |\delta x|
\end{equation}
This result is very interesting because equation \ref{average} implies that the speed of light possesses an average value but can fluctuate around it. And in particular, it could exceed the classical speed of light. This could lead to important phenomenology considerations and local tiny causality violations. However, \ref{average} does not tell us how many and how much those fluctuations appear.   

\section{Conclusion and Outlook}

We have investigated causal structure in noncommutative geometry. The analysis is based on an algebraization of causality, which can be implemented in the Lorentzian Spectral Triple.  
Space-time events are now described by pure states and causal evolution encoded with causal functions. All this formalism is then applied to $\kappa$-Minkowski space. Let us comment on this construction.\\
First, there is an infinite set of derivatives in $\kappa$-Minkowski; for the present construction of the Spectral Triple, one should only keep $\dd$ (i.e., the dimension of the space) of them. A earlier attempt in \cite{Franco_2022} was to consider derivatives $D_0 := \partial_0$ and $D_1 := x_1 \partial_1$, motivation of this choice was the study of $*$-automorphism group related to $\mathcal{A}$, see \cite{Franco_2022} for details. However, they do not have a suitable limit for the commutative regime. Here, we deal with this problem by choosing an appropriate derivative with a good limit and compatible with the axiom of the Lorentzian twisted Spectral Triple. \\
Another choice we made is the faithful $*$ representation given in the present case by the unitary representations of the affine group $\mathcal{G}$. This choice is far from being unique, and other $*$-representations, such as the GNS one, could provide a natural way to construct a twisted Lorentzian Spectral Triple for $\kappa$-Minkowski. \\
The impact of different constructions is still unclear, even if it \cite{Franco_2022}, they end up with formally similar quantum constraints for different Dirac operators. Such analog of the quantum speed of light could be seen as a feature of $\kappa$-Minkowski quantum causality in the formalism of Lorentzian Spectral Triple (with or without twist). \\
Regarding the quantum constraint \ref{average}, it has been derivate for a very particular kind of function, $a(x_0 , x_1) = x_0 \pm x_1$ and pure states \ref{pureS}. However, the set of pure states and causal functions solutions of \ref{causalconeexplicit} is not fully determined, and this \ref{average} is a particular quantum constraint among others. The entire characterization of the causal cone is still under investigation. Nevertheless a relevant family of functions, the light-cone coordinates $a(x_0,x_1) = x_0 + \lambda x_1, \quad \lambda \in [-1,1]$ belong to the causal cone. Those functions can be seen as describing an analog of the quantum causal cone that partly encodes causality. \\

\bibliographystyle{bib-style}
\nocite{*}
\bibliography{main}

\end{document}